\documentstyle[amssymb,12pt,thmsa,sw20lart]{article}

\input{tcilatex}
\begin{document}
\title{Selection rules at the quark-antiquark vertex of the QCD Pomeron
}

\author{H. Navelet and R. Peschanski\thanks{%
CEA, Service de Physique
Th\' eorique, CE-Saclay, F-91191 Gif-sur-Yvette Cedex, France}
}
\maketitle
\begin{abstract}
We derive the full analytic expression for the QCD eikonal coupling of a quark-antiquark state to the exchanged gluon-gluon state in the BFKL formalism. The formula is
valid for all conformal spin configurations of the $q\overline{q}$ and $gg$
states. In particular, a new selection rule on conformal spins characterizes the non-dominant BFKL components with intercept below the Pomeron in the conformal-invariant framework. 
\end{abstract}
\bigskip
{\bf 1.}
The new experimental results on deep-inelastic physics at small $x$ (HERA, Tevatron) have contributed to revive the old but
pertinent  QCD approach of Lipatov and collaborators \cite{BFKL}.
This approach allows one to calculate the resumed perturbative contribution of gluon-gluon
states to the hard QCD Pomeron. This contribution is expected to  dominate the structure functions
at small $x$ in the leading $\left( \alpha \,\log 1/x\right) ^{n}$ orders
(LLA approximation). However in all known physical situations, it remains to
determine the coupling of this gluon-gluon exchanged state to the $q%
\overline{q}$ state present in the projectile and/or target wave-function. It is  for instance the case in the QCD dipole model \cite{mueller}
 of the virtual
photon and the proton \cite{ourpap} interacting in deep-inelastic scattering.

\vskip 0.5cm

The problem of coupling the BFKL Pomeron to external quarks (or anti quarks)
has already been addressed  \cite{tang,bartels}. A first approach considers a local coupling to quark \cite{tang}, but it explicitely  spoils the conformal invariance of the
theory. As shown later on \cite{bartels}, conformal invariance can be preserved,
provided one uses a wave-function of the incident $q\overline{q}$ pair
obeying gauge invariance properties \cite{lip}. A convenient way of satisfying
these constraints is to consider the eikonal coupling of the $q\overline{q}$
state to the two-gluon exchanged state \cite{tang,bartels,forshaw}. In particular, the
initial dipole states described in the framework of the QCD dipole model \cite{ourpap}
satisfies this eikonal prescription.

\vskip 0.5cm

The aim of our paper is to derive the most general eikonal coupling of a $q\overline{q}$ state to the BFKL Pomeron. In a first part {\bf 2.} we recall the properties of the conformal-invariant basis giving a complete description of
the BFKL Pomeron states. In a second part {\bf 3.} we derive the most general
expression of the eikonal vertices in terms of this conformal  basis and explicitly compute the whole set of components labelled by the conformal spin $n$ of  the gluon-gluon state  and the conformal spin $n^{\prime}$ of the $q\overline{q}$ one ($\left(n,\ n^{\prime }\right)\ \in {\Bbb Z} $). We generalize the previous result \cite{forshaw}
obtained for $n=n^{\prime}=0,$ and, as shown in section {\bf 4.} find new selection rules, namely
 $n-n^{\prime }\equiv 0 \ (mod. 4).$ Interestingly enough, this leads to a first secondary conformal-invariant trajectory with vacuum quantum numbers and intercept around $\frac 12$
below the Pomeron. The solution is found to be a couple of convergent series in transverse momentum. Section {\bf 5.} summarizes our results. Appendix {\bf A1} gives  the
derivation of a {\it pseudo orthogonality} relation between conformal eigenvectors, which is our main technical tool.

\vskip 0.5cm

\noindent \textbf{2.\quad }We first recall the main results of Lipatov \cite{lip} for the scattering amplitudes of colorless objects in QCD in the
L.L.A.,
\begin{equation}
 A\left( s,t\right) =is\int \frac{d\omega }{2i\pi }s^{\omega}\ f_\omega\left( q^{2}\right);\ \ 
t=-q^{2}
\tag{1} 
\end{equation}
with 
\begin{equation}
f_\omega \left( q^{2}\right) =\int d^{2}k\ d^{2}k^{\prime}\ 
\phi ^{(1)}\left( k,q\right) \phi ^{(2)}\left( k^{\prime },q\right) f_\omega \left(
k,k^{\prime },q\right) 
\tag{2}
\end{equation}
where $f_\omega \left( k,k^{\prime },q\right) $ can
be interpreted as the $t$-channel partial-wave amplitude and $k^{\prime },k$
are the 2-dimensional transverse components of the exchanged
gluon momenta (see fig.1).
 The  vertex functions $\phi ^{(1),(2)}\left( k,q\right) $
characterize the internal structure of the colliding states and
can be calculated from perturbation theory in some cases. Our aim is to provide general rules obeyed by the vertex functions in a conformal invariant framework.

For this sake it is convenient to use the representation in terms of  impact parameters $\rho _i$%
\begin{eqnarray}
\delta ^{(2)}\left( q\!-\!q^{\prime }\right) f_\omega \left( k, k^{\prime
},q\right) =\left( 2\pi \right) ^{-8}\int \stackunder{r=1,2}{%
\prod }d^{2}\rho _r\stackunder{r^{\prime }=1,2}{\prod }d^{2}\rho _r^{\prime }
\times \nonumber \\
 f_\omega \left( \rho _{1},\rho _{2},\rho _{1}^{\prime },\rho _{2}^{\prime
}\right)\ \exp \left[ ik\rho _{1}+i\left( q\!-\!k\right) \rho _{2}-ik^{\prime
}\rho _{1}^{\prime }-i\left( q^{\prime }\!-\!k^{\prime }\right) \rho
_{2}^{\prime }\right] 
\tag{3} 
\end{eqnarray}
\noindent Following Ref.[6]:
\begin{eqnarray}
&&\delta ^{\left( 2\right) }\left( q-q^{\prime }\right) f_{\omega }\left(
q^{2}\right) = \frac{1}{2\pi^{8}}\int d^{2}\rho _{1}\;d
^{2}\rho _{2}\;d ^{2}\rho _{1}^{\prime }\;d ^{2}\rho _{2}^{\prime
}\ \hat{\phi}^{\left( 1\right) }\left( \rho _1,\rho _{2},q\right)\nonumber \\
&&\qquad \qquad \times \ \hat{\phi} ^{\left( 2\right) }\left( \rho
_{1}^{\prime }\rho _{2}^{\prime },q^{\prime }\right) f_{\omega }\left( \rho
_{1},\rho _{2},\rho _{1}^{\prime },\rho _{2}^{\prime }\right)
\tag{4}
\end{eqnarray}
where
\begin{equation}
\hat{\phi}\left( \rho _{1},\rho
_{2},q\right) = \int d^{2}k\ \phi \left( k,q\right)
\ e^{ik\rho _{1}}\ e^{i\left( q-k\right) \rho _{2}}. 
\tag{5}
\end{equation}
Note that, by virtue of gauge invariance
\begin{equation}
\phi ^{(1,2)}\left( k,q\right)\vert_{k=0}=\phi ^{(1,2)}\left( k,q\right)\vert_{k=q}\equiv 0.
\tag{6} 
\end{equation}
In the BFKL formalism, using the conformal invariant basis $E^{n,\nu}\left( \rho _{i\alpha },\rho _{j\alpha }\right),$ where $  \rho _{kl} \equiv \rho _{k} -
\rho _{l},$
one obtains
\begin{eqnarray}
f_\omega \left( \rho _{1},\rho _{2},\rho _{1}^{\prime },\rho _{2}^{\prime
}\right) &=&\stackunder{n}{\sum }\int \frac{c\left( n,\nu \right)\ d\nu }{
\left[\ \omega -\omega \left( n,\nu \right) \right] }\nonumber \\
&&\times
\ \int d^{2}\rho
_{0}\ E^{n,\nu }\left( \rho _{1^{\prime }0},\rho _{2^{\prime }0}\right)
E^{n,\nu }\left( \rho _{10},\rho _{20}\right),
\tag{7}
\end{eqnarray}
with
\begin{equation}
c(n,\nu) =\left( \nu ^{2}+\frac{n^{2}}{4}\right)\left\{\left[\nu^2\!+\!\left(\frac {n\!-\!1}2\right)^2\right]
\left[\nu^2\!+\!\left(\frac {n\!+\!1}2\right)^2\right]\right\}^{-1},
\tag{8}
\end{equation}
where $E^{n,\nu }\left( \rho _{i\alpha },\rho _{j\alpha }\right) $ are the $%
SL(2,{\Bbb C})$ eigenvectors corresponding to eigenvalues $\omega \left( n,\nu
\right) $ defined by the quantum numbers $\nu \in \Re,$ the conformal dimension and  $%
n \in {\Bbb Z},$ the conformal spin. 
\begin{equation}
 \omega \left( n,\nu \right) =\ \frac{2\alpha _{s}N_{c}}{\pi }
\left[ \psi\left( 1\right)
 -{\cal R}e \left\{\psi \left( \frac{1+\vert n\vert}2 +i\nu\right) \right\}\right]
\tag{9} 
\end{equation}
\begin{equation}
 E^{n,\nu }\left( \rho _{i\alpha },\rho _{j\alpha }\right) =\left( \frac{\rho
_{ij }}{\rho _{i\alpha }\rho _{j\alpha }}\right) ^{\mu +1/2}\left( 
\frac{\bar {\rho} _{ij }}{\bar{\rho}_{i\alpha }\bar{\rho}_{j\alpha }}\right) ^{
\tilde{\mu}+1/2}\left( -1\right) ^{\mu -\tilde{\mu}} 
\tag{10} 
\end{equation}
where
\begin{equation}
 \mu  =i\nu -\frac{n}{2}
;\  \tilde{\mu} = i\nu +\frac{
n}{2}.
\tag{11} 
\end{equation}
Now it is convenient to introduce the mixed representation of these
eigenvectors using a Fourier transform
\begin{equation}
E_{q}^{n,\nu }\left( \rho _{ij} \right) =\frac{2\pi ^{2}}{b_{n,\nu }}\ \int 
\frac{d^{2}\rho _0}{\left| \rho _{ij}\right| }\ e^{i q \left(\frac
{\rho _{i0} + \rho _{j0}}2\right)}
\ E^{n,\nu }\left(\rho _{i0},\rho _{j0}\right),
\tag{12} 
\end{equation}
\noindent where $b_{n,\nu }$ is given in Ref. [6].
Note that there exists an analytic expression \cite{conformal} for $E_{q}^{n,\nu }\left( \rho \right), $ namely
\begin{eqnarray} 
E_{q}^{n,\nu }\left( \rho \right) 
=\left( \frac{\bar{q}}{8}\right) ^{\mu
}\left( \frac{q}{8}\right) ^{\tilde{\mu}}\Gamma\left( 1/2-\mu \right)
\Gamma\left( 1/2-\tilde{\mu}\right) \nonumber \\ 
 \times \ \left[ J_{-\mu }\left( \frac{\bar{q}\rho }{4}\right) J_{-{\tilde{\mu}%
}}\left( \frac{q\bar{\rho}}{4}\right) -\left( -1\right) ^{\mu -\tilde{\mu}%
}J_{\mu }\left( \frac{q\bar{\rho}}{4}\right) J_{-{\tilde{\mu}}}\left( \frac{%
q\bar{\rho}}{4}\right) \right].
\tag{13} 
\end{eqnarray}
The completeness of the conformal basis implies an orthogonality relation
in the mixed representation \cite{conformal}
\begin{equation}
\frac{1}{4\pi ^{2}}\int \frac{d^{2}\rho }{\left| \rho \right| ^{2}}
E_{q}^{n,\nu }\left( \rho \right) \bar{E}_{q}^{n^{\prime },\nu ^{\prime
}}\left( \rho \right) =\delta _{n,n'}\delta \left( \nu \!\!-\!\!\nu ^{\prime
}\right) +\delta _{-n,n'}\delta \left( \nu \!\!+\!\!\nu ^{\prime }\right) \left(
q\right)^{2\tilde{\mu}}\left(
\bar {q}\right)^{2\mu}\ e^{i\delta(n,\nu)},
\tag{14}
\end{equation}
where the phase $e^{i\delta(n,\nu)}$ is defined in Ref.[6].

\bigskip
{\bf 3.}
 Using now the expression (4) in
the definition (3) of $f_\omega \left( q^{2}\right) $
 yields after some algebra 
\begin{eqnarray}
\delta ^{2}\left( q\!-\!q^{\prime }\right) f_\omega \left( q^{2}\right) =\left(
2\pi \right) ^{-8}\stackunder{n}{\sum }\int d^{2}\rho _1d^{2}\rho
_{2}d^{2}\rho _{1}^{\prime }d^{2}\rho _{2}^{\prime }\ d^{2}kd^{2}k^{\prime
} \nonumber\\ \times
\int \frac{c\left( n,\nu \right)\ d\nu} {
 \left( \omega -\omega \left( n,\nu \right) \right) } 
\int d^{2}\rho _{0}\ e^{i\left( q-q^{\prime }\right) \rho _{0}}\overline {E}^{n,\nu }\left( \rho _{1^{\prime }0,}\rho _{2^{\prime
}0}\right) E^{n,\nu }\left( \rho _{10},\rho _{20}\right) \nonumber \\ \times
\phi ^{(1)} \left( k,q\right) \phi ^{(2)}\left( k^{\prime },q\right) e^{\left[
ik\rho
_{10}+i\left( q-k\right) \rho _{20}-ik^{\prime }\rho _{1^{\prime }0}-i\left(
q^{\prime }-k^{\prime }\right) \rho _{2^{\prime }0}\right]}
\tag{15}
\end{eqnarray}
The integral over $d^{2}\rho _{0}$ yields the expected $2\pi ^{2}\ \delta ^{\left(
2\right) }\left( q-q^{\prime }\right) $
 and we get 
\begin{equation}
f_\omega \left( q^{2}\right) = \stackunder{n}{\sum }\int \frac{c\left(
n,\nu \right)\ d\nu } {\left( \omega -\omega \left( n,\nu \right) \right) }\ 
V_{1}^{n,\nu }\left( q\right) \ \overline{V}_{2}^{\ n,\nu }\left( q\right)\ ,
\tag{16} 
\end{equation}
\noindent with 
\begin{equation}
 V_{1}^{n,\nu }\left( q\right) =\frac {1}{\left( 2\pi \right) ^{3}}\ \int
d^{2}\rho _{10} d^{2}\rho _{20} d^{2}k\ \phi ^{(1)}\left( k,q\right) e^{i\left[
k\rho _{10}-\left( k-q\right) \rho _{20}\right] }\ E^{n,\nu }\left( \rho
_{10},\rho _{20}\right),
\tag{17} 
\end{equation}
and a similar expression for $V_{2}.$
Formula (16) clearly exhibits the  factorization between, respectively, the BFKL kernel, the upper, and the lower vertex of the
QCD t-channel partial waves.

Let us now discuss the functions $\phi ^{(1,2)}.$ From now on, we shall assume that the exchanged gluon-gluon state is linked to a quark-antiquark color
singlet, through an eikonal current. Under this physical assumption related to a well-known semi-classical description of scattering, the
functions $\phi ^{(1)}$ and $\phi ^{(2)}$ read 
\begin{equation}
\phi ^{(1)}\left( k,q\right) =\int d^{2}r\ f_{1}\left( r\right) 
\left\{ e^{i
\frac{k}{2}r}-e^{-i\frac{k}{2}r}\right\}\ 
\left\{ e^{i \frac{\left(
q-k\right)}{2}r}-e^{-i \frac{\left( q-k\right)}{2}r}\right\}
\tag{18}
\end{equation}
The eikonal formulation is such that   the QCD gauge invariance relations (6) are automatically fulfilled. $f_{1}\left( r\right) $ is the function
which describes the internal structure of the incident state and thus is momentum-independent.
In formula (17) for the vertex function $V_{1}^{n,\nu }\left( q\right) ,$ the integral over $d^{2}k$ can thus be
easily performed. Indeed 
\begin{eqnarray}
&&\frac 1 {\left( 2\pi \right)^2}\ \int d^{2}k\ e^{iq\rho _{20}}e^{ik\left( \rho _{10}-\rho _{20}\right) }\left[ \left( e^{i\frac{q}{2}r}+c.c.\right)
-\left( e^{-i\frac{q}{2}
r}e^{ikr }+c.c.\right) \right] \nonumber \\
&&= e^{iq\rho _{20}}\left( e^{iq\frac{r}{2}}+c.c.\right)
\delta ^{\left( 2\right) }\left( \rho _{10}-\rho _{20}\right) \nonumber \\
&& - e^{iq\rho
_{20}}\left[ e^{-i\frac{q}{2}r}\delta ^{\left( 2\right) }\left( \rho
_{10}-\rho _{20}+r\right)  +e^{i\frac{q}{2}r}\delta ^{2}\left( \rho _{10}-\rho _{20}-r\right)
\right]\ .
\tag{19}
\end{eqnarray}
\noindent The $\delta ^{2}\left( \rho _{10}-\rho _{20}\right) $ term gives
no contribution since
$ E^{n,\nu }\left( \rho _{10},\rho _{20}\right)$ vanishes at $\rho
_{10}=\rho _{20}.$

The last term in (19) yields after $\rho _{10}$-integration
\begin{eqnarray}
V_{1}^{n,\nu }\left( q\right) &=&-\int d^{2}r\ f_{1}(r)\int d^{2}\rho
_{20}\left\{ e^{iq\left( \rho _{20}\!+\!\frac{r}{2}\right) }E^{n,\nu }\left( \rho
_{20}\!+\!r,\rho _{20}\right) +(r \Rightarrow \!-\!r)\right\} \nonumber \\
&=&-2\int d^{2}r\ f_{1}\left( r\right) \int d^{2}R\ e^{iqR}E^{n,\nu }\left(
R-\frac{r}{2},R+\frac{r}{2}\right)\nonumber \\
&=&\frac{b_{n,\nu} }{\pi ^2}\int d^{2}r\ f_{1}\left(
r\right) \ \left| r\right| E_q^{n,\nu }\left( r\right).
\tag{20}
\end{eqnarray}

\noindent Note that using the eikonal coupling is nothing but projecting  
$f_{1}\left( r\right) $ on the eigenvectors $E_q^{n,\nu }$ of the mixed representation.

To proceed further it is convenient to expand a generic function $%
f_{1}\left( r\right) $ on the complete basis $E^{n,\nu }.$
Here  this function depends  on $r=\sigma _{12},$ where
 $\sigma _1$
(resp.  $\sigma _2$), is the quark (resp. antiquark) transverse coordinate.
\begin{equation} 
f_{1}\left( r\right) =\stackunder{n^{\prime }}{\sum }\int d\nu ^{\prime
}\int \frac{d^{2}\sigma _{0}}{\left| \sigma _{12}\right| ^{4}}\ f_{1}^{n^{\prime },\nu ^{\prime }}\ \bar{E}^{n^{\prime },\nu ^{\prime
}}\left( \sigma _{10},\sigma _{20}\right).
\tag{21} 
\end{equation}
\noindent where the coefficients $f^{n,\nu }$ do not depend on $\sigma _{0}$ since $f_{1}$
describes the internal structure of the incoming state independently of the reference coordinate $
\sigma _{0}.$
Now 
\begin{equation}
\int d^{2}\sigma _{0}\ \bar{E}^{n^{\prime },\nu ^{\prime }%
}\left( \sigma _{10},\sigma _{20}\right) =\ \left| \sigma _{12}\right|
\bar{E}_{q=0}^{n^{\prime },\nu ^{\prime }}\ \frac {\bar{b}_{n^{\prime },\nu ^{\prime }}}
{ 2\pi ^{2}} \ ,
\tag{22}
\end{equation}
with
\begin{equation}
E_q^{n^{\prime },\nu ^{\prime }}\left( \sigma _{12}\right)\vert_{q=0} =\left(
\sigma _{12}\right) ^{\mu ^{\prime }}\left( \bar{\sigma}_{12}\right) ^{%
\tilde{\mu}^{\prime }}, \ \mu ^{\prime }=i\nu ^{\prime }-\frac{n^{\prime }}{2},
\ %
\tilde{\mu}^{\prime }=i\nu ^{\prime }+\frac{n^{\prime }}{2}
\tag{23}
\end{equation}
which obeys  the orthogonality relation (14) for $q=0,$ namely
\begin{equation}
\frac{1}{2\pi ^{2}}\int \bar {E}_0^{n^{\prime },\nu ^{\prime }}\left( \sigma
_{12}\right) E_0^{n,\nu }\left( \sigma _{12}\right) \frac{d^{2}\sigma
_{12}}{\left| \sigma _{12}\right| ^{2}}=\delta _{n,n^{\prime }}\ \delta \left(
\nu \!-\!\nu ^{\prime }\right) .
\tag{24} 
\end{equation}
The coefficients $f_{1}^{n^{\prime },\nu ^{\prime }}$ are readily obtained
by inversion 
\begin{equation}
f_{1}^{n^{\prime },\nu ^{\prime }}=\frac 1{
\bar {b}_{n^{\prime },\nu ^{\prime }}}\int d^{2}\sigma _{12}\ f_1\left(
\sigma _{12}\right) E_{q=0}^{n^{\prime },\nu ^{\prime }}\left( \sigma
_{12}\right) \left| \sigma _{12}\right|.
\tag{25} 
\end{equation}

Inserting the analytic form (23), we get at once
\begin{equation}
f_{1}^{n^{\prime },\nu ^{\prime }}=\frac{1}{\bar{b}
_{n^{\prime },\nu ^{\prime }}}\ \int rdrd\theta \left| r\right| ^{2i\nu
^{\prime }+1}\ e^{-in^{\prime }\theta}\ f_{1}\left( \vec{r}\right) 
\tag{26} 
\end{equation}
\noindent Note that in the isotropic case $ f_{1}\left( \vec{r}\right)
 = f_{1}\left( r \right) .$ The angular integration yields $n^{\prime }=0$ and $f_{1}^{0,\nu
^{\prime }}$ is nothing but the Mellin transform of $f_{1}\left(
\left| r\right| \right).$

The final expression for the vertex function $V_{1}^{n,\nu }\left( q\right) $ reads
\begin{equation}
V_{1}^{n,\nu }\left( q\right) = 2 b_{n,\nu }\stackunder{n^{\prime }}{\sum }\int
d^{2}\nu ^{\prime } \ f_{1}^{n^{\prime },\nu ^{\prime }}\frac{1}{2\pi ^2 }\int \frac{d^{2}r}{%
\left| r\right| ^{2}}\ E_q^{n,\nu }\left( r\right) \ {\bar {E}_0^{n^{\prime },\nu ^{\prime }}\left( r\right) }.
\tag{27}
\end{equation}
\noindent All amounts to compute the {\it pseudo orthogonality} relations (i.e. for $E_q$ and $E_0$), namely
\begin{equation}
 {\cal I}=-\frac{1}{2\pi ^{2}}\int \frac{d^{2}r}{\left| r\right| ^{2}}E_q^{n,\nu
}\left( r\right){\bar {E}_0^{n^{\prime },\nu ^{\prime }}\left( r\right) }.
\tag{28} 
\end{equation}

Crucial for the vertex evaluation, the calculation of the {\it pseudo orthogonality} relation (27) is given in the appendix {\bf A1}.  First, by mere symmetry property,  $E_q^{l,\lambda }\left( \vert r \vert e^{i\phi}\right)$ $= (-1)^l E_q^{l,\lambda }\left( \vert r \vert e^{i(\phi + \pi)}\right) $ where $\phi$ is the $(r, q)$ angle, only the configurations with integer $\frac {n-n ^{\prime }}2$ do contribute, elsewhere ${\cal I} \equiv 0.$ In the latter case, the result reads
\begin{equation}
{\cal I}=\frac 1{4\pi} (-1)^{\frac {n-n ^{\prime }}2}\left[\frac q8\right]^{\tilde {\mu}-\tilde {\mu}^{\prime}}\ 
\left[\frac {\bar q}8\right]^{ {\mu}- {\mu}^{\prime}}\frac {\Gamma (1\!-\! \mu)}{\Gamma (\tilde \mu)}\ 
\frac{\Gamma \left(\frac { {\mu}+ {\mu}^{\prime}}2\right)
\Gamma \left(\frac {- {\mu}+ {\mu}^{\prime}}2\right)}
{\Gamma \left( 1\!-\!\frac {\tilde {\mu}+\tilde {\mu}^{\prime}}2\right)
\Gamma \left( 1\!-\!\frac{\tilde {\mu}^{\prime}-\tilde {\mu}}2\right)}.
\tag{29}
\end{equation} 

{\bf 4.} 
The result (29) leads to quite noticeable selection rules. Indeed, the integral $\cal I$ does exhibit simple poles which are only located at
\begin{equation}
{\mu}+ {\mu}^{\prime} = -2p_1\ ;\  {\mu}^{\prime}-{\mu} = -2p_2\ ;\ p_{1,2} \in {\Bbb N}.
\tag{30}
\end{equation} 
At once, using the definition (23) one notes that poles contributing to the $\nu ^{\prime }$ integral of the vertex (27) have the same imaginary part $i\nu ^{\prime } = \pm i \nu$ and, for the real part, $\frac {n\pm n ^{\prime }}4$ is an integer. In fact, it is already known \cite{conformal} that the BFKL Pomeron exchange implies $n$ even, from symmetry of the reaction. Hence, the relation boils down to $\frac {n - n ^{\prime }}4$ being an integer.

This result is meaningful.  The {\it pseudo orthogonality} relation (29) tells us that there exists a  stringent selection rule, since only spacing by 4 is allowed,  namely 
\begin{equation}
n - n^{\prime} = 0,\pm 4,\pm 8,...,\pm 4p \ ; p \in {\Bbb N}.
\tag{31}
\end{equation}
As a consequence, for an isotropic vertex $n^{\prime} = 0,$ only  $%
SL(2,{\Bbb C})$ eigenvectors with $n = 0 \cdot \cdot \cdot \pm 4p$ do contribute. Conversely, the $n=0$ component, which is the BFKL Pomeron
is coupled only to vertex coefficients with $n' = 0 \cdot \cdot \cdot \pm 4p.$ 

Conformal invariance thus predicts    subassymptotic contributions in energy, the first one being of the form (see e.g.
formula (1):
\begin{equation}
A(s,t=0) \simeq s^{\omega(n=4,\nu\simeq 0)}
,
\tag{32}  
\end{equation}
where the value $\nu\simeq 0$ corresponds to the dominant sadddle-point at $t=0$ and,  from (9), the intercept is given by,
\begin{equation}
 \omega \left( n=4,\nu =0\right) =\ \frac{2\alpha _{s}N_{c}}{\pi }\left[ \psi
\left( 1\right) -\psi \left(5/2\right) \right]\ .
\tag{33} 
\end{equation}
The high-energy behaviour (32) has to be compared with the dominant BFKL behaviour given by $s^{\omega \left( 0,0\right)}.$  The difference of intercepts is thus
\begin{equation}
 \omega \left( 0,0\right)- \omega \left( 4,0\right) =
\frac{2\alpha _{s}N_{c}}{\pi } \left[ \psi
\left( 1/2\right) -\psi \left(5/2\right) \right]
= \frac {16}3
\frac{\alpha _{s}N_{c}}{\pi }.
\tag{34} 
\end{equation}
Interestingly enough, a phenomenological determination of the BFKL intercept
based on proton structure functions gives $\omega \left( 0,0\right) \simeq .3 \cite {ourpap}$ and  leads to $\omega \left(0,0\right) - \omega \left(4,0\right) = \frac 3{4\ln 2}\ \omega \left(0,0\right) \simeq 1.9 \ \omega \left(0,0\right)$ of the order $0.6.$  It is thus likely  that conformal invariance selection rules does not contradict some aspects of subassymptotic corrections to the Pomeron.

Using the {\it pseudo-orthogonality} relation (29), one finally gets the expression of the vertex functions, namely
\begin{eqnarray}
V_{1}^{n,\nu }\left( q\right) &=& \frac {b_{n,\nu }}{2\pi} (-1)^{\frac {n-n ^{\prime }}2}  \  \frac {\Gamma (1\!-\! \mu)}{\Gamma (\tilde \mu)}\ 
\stackunder{n^{\prime }}{\sum }\int
d^{2}\nu ^{\prime } \ f_{1}^{n^{\prime },\nu ^{\prime }} \nonumber \\  &\times&
\left[\frac q8\right]^{\tilde {\mu}-\tilde {\mu}^{\prime}}\left[\frac {\bar q}8\right]^{ {\mu}- {\mu}^{\prime}} \frac{\Gamma \left(\frac { {\mu}+ {\mu}^{\prime}}2\right)
\Gamma \left(\frac {- {\mu}+ {\mu}^{\prime}}2\right)}
{\Gamma \left( 1-\frac{\tilde {\mu}+\tilde {\mu}^{\prime}}2\right)
\Gamma \left(1-\frac {-\tilde {\mu}+\tilde {\mu}^{\prime}}2\right)}.
\tag{35}
\end{eqnarray}

In order to take advantage of the simple pole structure of $\cal I,$ we have to deform the imaginary $z=i\nu^{\prime}$ integration contour upon the real axis. In order to do so, we have to discuss both the convergence properties of $\cal I$ for large $z$ modulus and the  analytical structure and  convergence properties of the vertex coefficients $f_{1}^{n^{\prime },-iz}.$ We obtain for large
$\vert z\vert :$ 
\begin{equation}
\vert {\cal I}\vert \ \simeq
\  \left[\frac {q{\bar q}}{64}\right]^{-\Re z}\ \Gamma ^4(\frac z2)\ \propto
\left\{\frac {\vert z \vert}2\right\}^{-2 \Re z}\ .
\tag{36}
\end{equation} 
It is clear from (36) that, in absence of a modification of the dominant behaviour of the vertex integrand by the vertex coefficients $f_{1}^{n^{\prime },-iz},$
the integral contour may be closed on the left of the $z$-plane, picking the two series of pole contributions at $z= i\nu -\frac {n-n'+4p_1}2 $ and $z= -i\nu +\frac {n+n'-4p_2}2,$ provided  $\Re z $ be a negative integer. This gives two  convergent series for all values of $q,$  whose contributions are obtained at the poles fixed by  the BFKL dynamics.

Conversely, if the convergence properties of the vertex coefficients $f_{1}^{n^{\prime },-iz}$ allow a contour deformation to the right of the complex $z$-plane,  one obtains another convergent series for all values of $q,$ but picking up now the singularities of these vertex coefficients. 

An intermediate and interesting alternative
is when there is a matching of the large $z$ behaviour of ${\cal I}$ and $f_{1}^{n^{\prime },-iz}.$ In this case  one may deform the contour to the left (resp. to the right),  if $q<q_c,$ (resp. $q>q_c$), where the radius of convergence  $q_c$ is easily determined knowing the behaviour of $f_{1}^{n^{\prime },-iz},$ compared to that of ${\cal I}$. 

Note that in all three cases, one obtains a finite answer for every value of the momentum transfer $q.$ In the first two cases, there is a unique regime for the $q$ dependence of the vertex function, which is determined either by BFKL dynamics or by the vertex coefficients. In the third (intermediate) configuration, one observes the existence of two regimes, one for small $q$ with BFKL dynamics and one at large  $q$ with vertex dynamics. As an illustration of this discussion, let us consider some physical examples \cite{forshaw} of the function $f_1(r),$
corresponding to the simple isotropic case. A typical hadron non-perturbative probability distribution may be of the form   
$f(r) \propto e^{-Q_h^2 r^2},$ where $Q_h$ is a typical (small) hadronic scale. Using formula (26), this leads to $f_1^{0,-iz} \propto \Gamma(-z)$ at large z.
Hence, this belongs to the first case (left contour). For a high-$Q^2$ virtual  photon external state, perturbative QCD leads to $f(r) \propto K_{0,1}(Qr),$ and thus to $f_1^{0,-iz} \propto \Gamma^2(-z) \simeq \Gamma^4(-\frac z2)$ at large z, so that we are in the third case with a finite radius of convergence. This is the case with two different regimes of vertex $q$-dependence. Notice that, for a non-perturbative type of coupling our conclusions differ from \cite{forshaw},
since in this case, we have shown that there is only one, BFKL-dominated, regime
instead of two.

{\bf 5.} 
Let us summarize our main results on the general eikonale coupling of a 
quark-antiquark state to the BFKL Pomeron:

{\bf i)}
The conformal invariance properties of the BFKL kernel lead to a general exact solution for the vertex functions using the  $%
SL(2,{\Bbb C})$ expansion of the quark-antiquark probability distribution. It amounts to the projection of this probability distribution on the 
 $%
SL(2,{\Bbb C})$ eigenvector $E_q^{n,\nu },$ see formulae (20,33), where use is made of a {\it pseudo orthogonality} relation involving  $E_q^{n,\nu }$
and $E_0^{n,\nu },$ where $q$ is the transverse momentum of the reaction.

{\bf ii)}
Non-trivial selection rules are obtained for the conformal spins $n$ and $n',$
where the former governs the energy dependence of the Pomeron components and the latter is associated with vertex function components. One gets $n- n' =4p,$
with $p$ integer. 

{\bf iii)}
As a consequence, an isotropic vertex ($n'=0$) leads to a selection rule on the subassymptotic components of the BFKL kernel. Their effective difference of intercept with the Pomeron is given by
\begin{equation}
 \omega \left( 0,0\right)- \omega \left( 4p,0\right) =
\frac{2\alpha _{s}N_{c}}{\pi } \left[ \psi
\left( 1/2\right) -\psi \left(\frac {4p+1}2\right) \right].
\nonumber 
\end{equation}
Phenomenologically, the first subassymptotic component is around $.6$ below the BFKL Pomeron, and thus could play a r\^ ole for models using the BFKL formalism. Conversely, the vertex function components coupled to the main BFKL component ($n=0$)
are limited to conformal spins $n'=4p,$ which may indicate interesting selection rules on the angular momenta of the $q\bar q$ states coupled to the QCD Pomeron.

{\bf iv)}
The solution for the vertex can be expressed as a convergent series for all values of $q.$ More precisely, In almost all cases, the series is given by the 
residues of the poles of the BFKL coefficient function $\cal I,$ see formula (29). In some cases, e.g. virtual photon QCD vertex, there is a finite radius of convergence $q=q_c,$ below which the same is true, while for $q>q_c$ the expansion is given by the residues of the poles related to the quark-antiquark probability distribution in the photon. In this case two regimes are present in the transverse momentum distribution.
\bigskip

{\bf Acknowledgments}

We thank Andrzej Bialas and Misha Ryskin for  stimulating discussions.
\eject

{\bf APPENDIX A1. Pseudo-orthogonality relations}
\bigskip
\bigskip

Using formulae (10-12) the integral $\cal I$ is expressed by
\begin{equation}
{\cal I}= (-1)^n \frac {b_{n,\nu}}{2\pi^2} \int d^2\rho d^2R e^{iq\cdot R}
\rho^{\mu + \mu '} \bar {\rho}^{\tilde \mu + \tilde \mu '}
\left[R^2-\frac {\rho^2}4\right]^{-\left(\mu +\frac 12\right)}
\left[\bar R^2-\frac {\bar \rho^2}4\right]^{-\left(\tilde \mu +\frac 12\right)},
\tag{A1}
\end{equation}

for $n-n ^{\prime }$ even, and is 0 for $n-n ^{\prime }$ odd.
The change of variable $\rho^2 = 4R^2 r$ yields
\begin{eqnarray}
{\cal I}&=& (-1)^n \frac {b_{n,\nu}}{\pi^2} \int d^2r 
r^{\frac {\mu + \mu '}2 -1} \bar {r}^{\frac {\tilde \mu + \tilde \mu '}2 -1}
\left[1-r\right]^{-\left(\mu +\frac 12\right)}
\left[1-\bar r\right]^{-\left(\tilde \mu +\frac 12\right)}
\nonumber \\
 &\times& \  2^{ \left(\mu + \mu '+ \tilde \mu + \tilde \mu '\right)   } \int d^2R \ e^{iq\cdot R}\ 
R^{\mu - \mu '-1} \ \bar {R}^{\tilde \mu - \tilde \mu '-1}.
\tag{A2}
\end{eqnarray}

Using now known mathematical identities \cite{lip,kawai,geronimo},
one writes
\begin{eqnarray}
 \int d^2r \ 
r^{\left(\frac {\mu + \mu '}2 -1\right)} \bar {r}^{\left(\frac {\tilde \mu + \tilde \mu '}2 -1\right)}
\left[1-r\right]^{-\left(\mu +\frac 12\right)}
\left[1-\bar r\right]^{-\left(\tilde \mu +\frac 12\right)} =
\nonumber \\
\frac {\sin \frac {\pi}2 \left(\mu + \mu '\right)\ \sin \frac {\pi}2 \left(\!-\!\mu + 1/2\right)}{\sin \frac {\pi}2 \left( \mu '\!-\!\mu + 1\right)}
\nonumber\\
\frac {\Gamma  \left(\frac {\mu + \mu '}2\right)\ \Gamma  \left(-\mu + 1/2\right)}{\Gamma  \left( \frac {\mu '\!-\!\mu + 1}2\right)}
\ \frac {\Gamma  \left(\frac {\tilde\mu + \tilde\mu '}2\right)\ \Gamma  \left(\!-\!\tilde\mu + 1/2\right)}{\Gamma  \left( \frac {\tilde\mu '\!-\!\tilde\mu + 1}2\right)},
\tag{A3}
\end{eqnarray}
and
\begin{eqnarray}
 \int d^2R e^{iq\cdot R}
R^{\mu - \mu '-1} \bar {R}^{\tilde \mu - \tilde \mu '-1}=
\left[\frac q2\right]^{\tilde \mu - \tilde \mu '}\ 
\left[\frac {\bar q}2\right]^{\mu - \mu '}\
\nonumber\\
e^{i\frac {\pi}2  \left(\mu '-\mu + \tilde \mu - \tilde \mu '\right)}
\sin \pi \left( \tilde \mu '-\tilde \mu \right)
\Gamma  \left(\mu '-\mu \right)\ \Gamma  \left(\tilde \mu '-\tilde \mu \right)
.
\tag{A4}
\end{eqnarray}
Using the $\Gamma$ doubling formula $\frac {\Gamma  \left(\mu '- \mu \right)}
{\Gamma  \left( \frac {\mu '- \mu + 1}2\right)}\equiv \frac {2^{\mu '- \mu - 1}}{\sqrt \pi}\ \Gamma  \left(\frac {\mu '-\mu}2 \right)$ and the definition \cite{lip} of $b_{n,\nu},$ one gets at last the equation (29) of the
text. namely
\begin{equation}
{\cal I}=\frac 1{4\pi} (-1)^{\frac {n-n ^{\prime }}2}\left[\frac q8\right]^{\tilde {\mu}-\tilde {\mu}^{\prime}}\ 
\left[\frac {\bar q}8\right]^{ {\mu}- {\mu}^{\prime}}\frac {\Gamma (1\!-\! \mu)}{\Gamma (\tilde \mu)}\ 
\frac{\Gamma \left(\frac { {\mu}+ {\mu}^{\prime}}2\right)
\Gamma \left(\frac {- {\mu}+ {\mu}^{\prime}}2\right)}
{\Gamma \left( 1\!-\!\frac {\tilde {\mu}+\tilde {\mu}^{\prime}}2\right)
\Gamma \left( 1\!-\!\frac{\tilde {\mu}^{\prime}-\tilde {\mu}}2\right)}. 
\tag{A5}
\end{equation} 
\eject

\eject

\end{document}